\RequirePackage{lineno}
\documentclass[onecolumn,aps,prc,showpacs,superscriptaddress,preprintnumbers,floatfix,nofootinbib]{revtex4}
\usepackage{epsfig,graphics}
\usepackage{graphicx}% Include figure files
\usepackage{dcolumn}% Align table columns on decimal point
\usepackage{bm}% bold math
\usepackage{amsmath}
\usepackage[usenames]{color}
\usepackage{lineno}
\usepackage{ulem} %% for strike-through
\usepackage[colorlinks,citecolor=blue]{hyperref}

\begin{document}

\title{Sensitivity analysis of the chiral magnetic effect observables using a multiphase transport model}

\author{Ling Huang}
\affiliation{Shanghai Institute of Applied Physics, Chinese Academy of Sciences, Shanghai 201800, China}
\affiliation{University of Chinese Academy of Sciences, Beijing 100049, China}
\affiliation{Key Laboratory of Nuclear Physics and Ion-beam Application (MOE), Institute of Modern Physics, Fudan University, Shanghai 200433, China}

\author{Mao-Wu Nie}
\affiliation{Institute of Frontier and Interdisciplinary Science, Shandong University, Qingdao, 266237, China}

\author{Guo-Liang Ma}
\email[]{glma@fudan.edu.cn}
\affiliation{Key Laboratory of Nuclear Physics and Ion-beam Application (MOE), Institute of Modern Physics, Fudan University, Shanghai 200433, China}
\affiliation{Shanghai Institute of Applied Physics, Chinese Academy of Sciences, Shanghai 201800, China}

%\date{\today}

\begin{abstract}
Because the traditional observable of charge-dependent azimuthal correlator $\gamma$ contains both contributions from the chiral magnetic effect (CME) and its background, a new observable of $R_{\Psi_{m}}$ has been recently proposed which is expected to be able to distinguish the CME from the background. In this study, we apply two methods to calculate $R_{\Psi_{m}}$ using a multiphase transport model without or with introducing a percentage of CME-induced charge separation. We demonstrate that the shape of final $R_{\Psi_{2}}$ distribution is flat for the case without the CME, but concave for that with an amount of the CME, because the initial CME signal survives from strong final state interactions. By comparing the responses of $R_{\Psi_{2}}$ and $\gamma$ to the strength of the initial CME, we observe that two observables show different nonlinear sensitivities to the CME. We find that the shape of $R_{\Psi_{2}}$ has an advantage in measuring a small amount of the CME, although it requires large event statistics. 

\end{abstract}


\maketitle

\section{Introduction}
\label{sec:intro}
Relativistic heavy-ion collisions provide us an unique way to explore the natures of quark gluon plasma(QGP) experimentaly~\cite{Adams:2005dq,Adcox:2004mh}. In order to probe the QGP, many observables have been carried out experimentally, such as jet quenching~\cite{Wang:1991xy,Adams:2003kv,Aamodt:2010jd} and collective flow~\cite{Ollitrault:1992bk,Romatschke:2007mq,Heinz:2013th,Xu:2007jv}. Recently, the chiral magnetic effect (CME) has been proposed as a good observable which reveals some topological and electromagnetic properties of the QGP.  In the early stage of relativistic heavy-ion collisions, an extremely large magnetic field can be created which can induce an electric current along the strong magnetic field for chirality imbalanced domains with a nonzero topological charge inside the QGP, i.e. chiral magnetic effect~\cite{Kharzeev:2004ey,Kharzeev:2007tn,Kharzeev:2007jp,Fukushima:2008xe,Fukushima:2009ft}.  The transitional observable to detect the CME is a charge-dependent azimuthal correlator, $\gamma=\langle cos(\phi_{\alpha}+\phi_{\beta} - 2\Psi_{RP})\rangle$, which has been widely investigated both experimentally and theoretically~\cite{Abelev:2009ac,Abelev:2009ad,Adamczyk:2013hsi,Adamczyk:2013kcb,Adamczyk:2014mzf,Abelev:2012pa,Khachatryan:2016got}. Unfortunately, the observable can not distinguish the CME signal from the large background clearly~\cite{Schlichting:2010qia,Pratt:2010zn,Bzdak:2009fc,Bzdak:2010fd,Liao:2010nv,Wang:2009kd,Bzdak:2012ia,Zhao:2018skm}, because many kinds of backgrounds can contribute to $\gamma$~\cite{Bzdak:2010fd,Wang:2009kd}. Recently, a new observable, namely the shape of $R_{\Psi_{m}}$, has been proposed to be a more sensitive probe to search for the CME signal. Many studies of the $R_{\Psi_{m}}$ observable have been reported~\cite{Ajitanand:2010rc,Magdy:2017yje,Bozek:2017plp,Feng:2018chm}. For examples, some studies show that the shape of $R_{\Psi_{m}}$ dustribution is convex due to background but concave due to the CME~\cite{Magdy:2017yje,Feng:2018chm}, but another study shows that $R_{\Psi_{m}}$ could be also concave due to the background only~\cite{Bozek:2017plp}. On the other hand, because the lifetime of magnetic field may be quite short due to the limited conductivity of QGP~\cite{Voronyuk:2011jd,Cassing:2013iz,Ding:2015ona}, it is questionable whether the CME signal formed in the early stage can survive from strong final state interactions since relativistic heavy-ion collisions actually involves many final dynamic evolution stages. It has been found out that a multiphase transport model(AMPT) is a good way to study the interplay between the CME and final state interactions in relativistic heavy-ion collisions~\cite{Ma:2011uma,Shou:2014zsa,Huang:2017pzx}. Ma et al.~\cite{Ma:2011uma} domenstrated that a 10\% initial charge separation due to the CME can describe experiment data of $\gamma$ correlator in Au+Au collisions at 200GeV, and only 1-2\% percentage of charge separation can remain finally due to strong final state interactions. In this study, we investigate the new observable of $R_{\Psi_{m}}$  with two settings of the AMPT models, the original AMPT model which contains backgrounds only and the AMPT model with introducing a CME-induced charge separation. We compare the shapes of $R_{\Psi_{m}}$ distributions from the background case and the CME case. We also study the relationship between strength of the CME between $R_{\Psi_{2}}$ and $\gamma$ in order to reveal the sensitivity of those observables to the CME.

This paper is organized as follows. We will introduce our methods of calculating $R_{\Psi_{m}}$ and how to introduce a CME-induced charge separation into the AMPT model in Section II. Our results and discussion are presented in Section III. 
\section{Model and calculation method}
\subsection{The AMPT model}
\label{sec:partA}
A multiphase transport model, AMPT, has been extensively used to investigate the physics of relativistic heavy-ion collisions~\cite{Lin:2004en,Ma:2016fve,Ma:2013gga,Ma:2013uqa,Bzdak:2014dia,Nie:2018xog}. In order to study the $R_{\Psi_{m}}$, we simulated Au+Au 200 collisions at 200 GeV (3mb) with the new version of AMPT model with string meting mechanism in which charges are strictly conserved. There are four main stages in the AMPT model~\cite{Lin:2004en}: the initial condition, partonic interactions, conversion from partonic to the hadronic matter and hadronic interactions. The initial condition mainly simulates the spatial and momentum distributions of minijet partons from QCD hard processes and soft string excitations by using HIJING model~\cite{Wang:1991hta,Gyulassy:1994ew}. The parton cascade describes strong interactions among partons through elastic partonic collisions only which are controlled by a partonic interaction cross section~\cite{Zhang:1997ej}. When all partons stop to interact, the AMPT model simulates hadronization by coalescence, i.e. comparing two nearest partons into a meson and three nearest quarks into a baryon. Finally, the ART model is used to simulate baryon-baryon, baryon-meson and meson-meson interactions~\cite{Li:1995pra}.  There is no the chiral magnetic effect in the original AMPT, so we need to add an additional CME-induced charge separation into the initial condition of the AMPT model in order to study CME-related physics. In previous work~\cite{Ma:2011uma}, the CME signal has been successfully introduced into the AMPT model by switching the $p_y$ values of a percentage of the downward moving $u$ ($\bar{d}$) quarks with those of the upward moving $\bar{u}$ ($d$) quarks to thus produce a charge dipole separation in the initial condition. In our convention, we always choose $x$ axis along the direction of impact parameter $b$ from the target center to the projectile center, $z$ axis along the beam direction, and $y$ axis perpendicular to the $x$ and $z$ directions. The percentage of initial charge separation is used to adjust strength of the CME. 
The percentage $f$ is defined as,
\begin{equation}
f = \frac{N_{\uparrow(\downarrow)}^{+(-)}-N_{\downarrow(\uparrow)}^{+(-)}}{N_{\uparrow(\downarrow)}^{+(-)}+N_{\downarrow(\uparrow)}^{+(-)}},
 \label{eq-f}
\end{equation}
where $N$ is the number of a given species of quarks, $+$ and $-$ denote positive and negative charges, respectively, and $\uparrow$ and $\downarrow$ represent the moving directions along the $y$ axis. Note that the relation between our $f$ and the usual $a_{1}$ is $f=(4/\pi)a_{1}$, where $a_{1}$ is the coefficient of $\mathrm{sin}\phi$ term in the Fourier expansion of particle azimuthal angle distribution. By taking advantage of two settings of AMPT model, i.e. without and with introducing the CME, we next will apply the new observable $R_{\Psi_{m}}(\Delta S)$ to systemically investigate how the new observable works for searching for the CME.

\subsection{Calculation methods}
\label{sec:partB}
Two methods, mixing-particle method~\cite{Ajitanand:2010rc} and shuffling-particle method~\cite{Magdy:2017yje}, are used to calculate the new observable of $R_{\Psi_{m}}$  for Au+Au collisions at 200GeV (30-50\%). Because the definition of $R_{\Psi_{m}}$ is based on another observable of $C_{\Psi_{m}}$, we firstly show the formulas for calculating $C_{\Psi_{m}}$ in the mixing-particle method as follows~\cite{Ajitanand:2010rc},
\begin{equation}\label{Rm1}
\langle S_{p^{+}} \rangle=\frac{1}{N_{p}}\sum_{1}^{N_p} \mathrm{sin}(\frac{m}{2}(\phi^{+}_{p}-\Psi_{m})),
\end{equation}
\begin{equation}
\langle S_{n^{-}} \rangle=\frac{1}{N_{n}}\sum_{1}^{N_n} \mathrm{sin}(\frac{m}{2}(\phi^{-}_{n}-\Psi_{m})),
\end{equation}
\begin{equation}\label{Rm2}
\Delta S=\langle S_{p^{+}}\rangle-\langle S_{n^{-}}\rangle,
\end{equation}
where $\phi$ is the azimuthal angle of particle, $\Psi_{m}$ is the event reaction plane,  superscript $+$ and $-$ sign particles' charges, $N_{p}$ and $N_{n}$ represent the total number of positive and negative charged particles, respectively. For $m$=2, the distribution of $\Delta S$ is expected to be broaden due to the existence of the CME.

In mixing-particle method, to make a corresponding reference of $\Delta S$, which is denoted as $\Delta S_{mix}$, we select the same number of particles as for $\Delta S$ but ignore their charges, and we can do similar calculations as follows,
\begin{equation}\label{Rm3}
\langle S_{p^{mix}}\rangle=\frac{1}{N_{p}}\sum_{1}^{N_p} \mathrm{sin}(\frac{m}{2}(\phi_p^{mix}-\Psi_{m}))
\end{equation}
\begin{equation}
\langle S_{n^{mix}}\rangle=\frac{1}{N_{n}}\sum_{1}^{N_n} \mathrm{sin}(\frac{m}{2}(\phi_n^{mix}-\Psi_{m}))
\end{equation}
\begin{equation}\label{Rm4}
\Delta S_{mix}=\langle S_{p^{mix}}\rangle-\langle S_{n^{mix}}\rangle.
\end{equation}
where we use superscript "mix" to sign mixing particles' charges. Then we can get $C_{\Psi_{m}}$ by taking the ratio of the distribution of $\Delta S$ [$N(\Delta S)$] and the distribution of $\Delta S_{mix}$ [$N(\Delta S_{mix})$].
\begin{equation}\label{Rm5}
C_{\Psi_{m}}(\Delta S)=N(\Delta S)/N(\Delta S_{mix}),m=2,3
\end{equation}

On the other hand, by shifting the $\Psi_{m}$ to $\Psi_{m}$+$\pi/m$, $C^{\perp}_{\Psi_{m}}(\Delta S)$ is expected to only reflect the background of the CME. We replace $\Psi_{m}$ with $\Psi_{m}$+$\pi/m$ in the above formulas, $C^{\perp}_{\Psi_{m}}(\Delta S)$ can be obtained as follows,
\begin{equation}\label{Rm6}
\langle S^{\perp}_{p^{+}}\rangle=\frac{1}{N_{p}}\sum_{1}^{N_p} \mathrm{sin}(\frac{m}{2}(\phi^{+}_{p}-\Psi_{m}-\frac{\pi}{m}))
\end{equation}
\begin{equation}
\langle S^{\perp}_{n^{-}}\rangle=\frac{1}{N_{n}}\sum_{1}^{N_n} \mathrm{sin}(\frac{m}{2}(\phi^{-}_{n}-\Psi_{m}-\frac{\pi}{m}))
\end{equation}
\begin{equation}\label{Rm7}
\Delta S^{\perp}=\langle S^{\perp}_{p^{+}}\rangle-\langle S^{\perp}_{n^{-}}\rangle
\end{equation}
\begin{equation}\label{Rm8}
\langle S^{\perp}_{p^{mix}}\rangle=\frac{1}{N_{p}}\sum_{1}^{N_p} \mathrm{sin}(\frac{m}{2}(\phi_p^{mix}-\Psi_{m}-\frac{\pi}{m}))
\end{equation}
\begin{equation}
\langle S^{\perp}_{n^{mix}}\rangle=\frac{1}{N_{n}}\sum_{1}^{N_n} \mathrm{sin}(\frac{m}{2}(\phi_n^{mix}-\Psi_{m}-\frac{\pi}{m}))
\end{equation}
\begin{equation}\label{Rm9}
\Delta S^{\perp}_{mix}=\langle S^{\perp}_{p^{mix}}\rangle-\langle S^{\perp}_{n^{mix}}\rangle
\end{equation}
\begin{equation}\label{Rm10}
C^{\perp}_{\Psi_{m}}(\Delta S)=N(\Delta S^{\perp})/N(\Delta S^{\perp}_{mix}),m=2,3.
\end{equation}

In the other method of shuffling-particle method, its formulas are same with those of mixing-particle method except for the definitions of $\Delta S_{mix}$ and $\Delta S^{\perp}_{mix}$. In the above mixing-particle method, $\Delta S_{mix}$ and $\Delta S^{\perp}_{mix}$ are obtained by ignoring charges when mixing all particles. But in shuffling-particle method, they are obtained by reshuffling their charges of charged particles, denoted as $\Delta S_{shuffle}$ and $\Delta S^{\perp}_{shuffle}$.

For both methods, once we get $C_{\Psi_{m}}(\Delta S)$ and $C^{\perp}_{\Psi_{m}}(\Delta S)$, $R_{\Psi_{m}}(\Delta S)$~\cite{Magdy:2017yje,Bozek:2017plp,Feng:2018chm} is obtained as,
\begin{equation}\label{Rm11}
R_{\Psi_{m}}(\Delta S)=C_{\Psi_{m}}(\Delta S)/C^{\perp}_{\Psi_{m}}(\Delta S).
\end{equation}
The shape of $R_{\Psi_{m}}(\Delta S)$ is expected to be sensitive to whether the CME exists or not. In our work, we will calculate $R_{\Psi_{m}}(\Delta S)$ with the two methods with the AMPT model without and with introducing a CME-induced charge speration, and the detailed results will be presented in the section III.

\section{Results and Discussions}
\label{sec:results}
In this work, we selected particles with transverse momenta 0.35$<$ $p_{T}$ $<$2.0 GeV/c and pseudorapidity -1.0$<$ $\eta$ $<$1.0 to calculate $C_{\Psi_{m}}$, $C^{\perp}_{\Psi_{m}}$ and $R_{\Psi_{m}}$. As for $\Psi_{m}$, the information of coordinate space in the initial stage are used for its reconstruction~\cite{Ma:2010dv}. Two methods are both applied for caculating $R_{\Psi_{m}}$. The results are presented in subsection IIIA. In order to investigate the relationship between R and the CME strength, the dependence of the CME observables on initial charge separation percentage have been also calculated, which is presented subsection IIIB.

\subsection{$C_{\Psi_{2}}$, $C^{\perp}_{\Psi_{2}}$ and $R_{\Psi_{2}}$}
\label{sec:part C}

\begin{figure}
\includegraphics[scale=0.85]{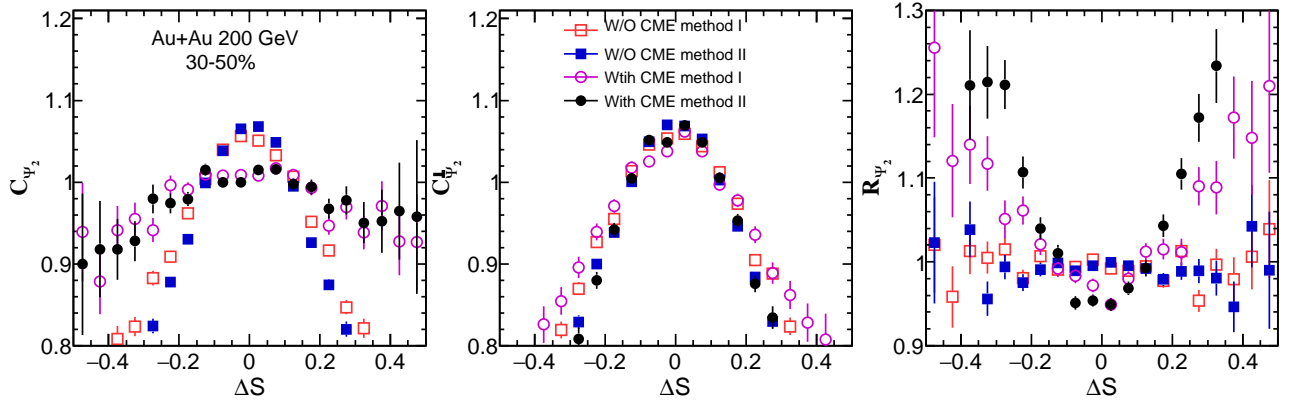}
\caption{(Color online)  $C_{\Psi_{2}}$, $C^{\perp}_{\Psi_{2}}$ and $R_{\Psi_{2}}$ in Au+Au collisions at 200GeV (30-50\%) from the AMPT model without or with the CME based on two different methods, where method I  and method II represent the mixing-particle method and the shuffling-particle method, respectively.
}
\label{fig-1}
\end{figure}

Since the original AMPT model does not include the CME, we can calculate $R_{\Psi_{2}}$ through it to study the pure background effect. On the other hand,  $R_{\Psi_{2}}$ from the AMPT model with introducing the CME can help us find the CME signal from the background. The results are presented in Fig.~\ref{fig-1},
which shows $C_{\Psi_{2}}$, $C^{\perp}_{\Psi_{2}}$ and $R_{\Psi_{2}}$ from the AMPT model without or with introducing an initial CME-induced charge separation based on two methods, where Method I denotes the mixing-particle method and method II denotes the shuffling-particle method. We found that the results from the two methods are consistent with each other. In their shapes, $C_{\Psi_{2}}$ and $C^{\perp}_{\Psi_{2}}$ are convex for original AMPT model without the CME, $R_{\Psi_{2}}$ is flat. On the other hand, $C_{\Psi_{2}}$ and $C^{\perp}_{\Psi_{2}}$ are convex  for the AMPT model with introducing a 10\% of CME-induced initial charge separation, but they are broadened differently due to the CME which makes the shape of $R_{\Psi_{2}}$ concave finally. From all curves in Fig.~\ref{fig-1}, $C_{\Psi_{2}}$ and $C^{\perp}_{\Psi_{2}}$ are convex no matter whether there is the CME or not. However, $R_{\Psi_{2}}$ is flat if with background only, but it becomes concave if introducing a 10\% of initial CME-induced charge separation.

\begin{figure}
\includegraphics[scale=0.9]{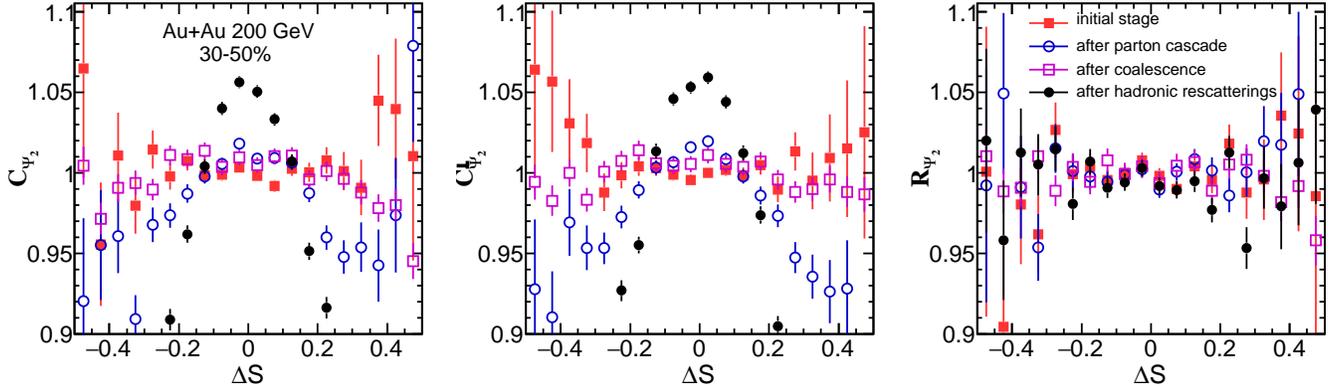}
\caption{(Color online) $C_{\Psi_{2}}$, $C^{\perp}_{\Psi_{2}}$ and $R_{\Psi_{2}}$ in Au+Au collisions at 200 GeV (30-50\%) for different evolution stages of the original AMPT model without the CME.
}
\label{fig-2}
\end{figure}

From the results in Fig.~\ref{fig-1}, we can see $R_{\Psi_{2}}$ can be a probe to distinguish the CME signal from the background. To understand why $R_{\Psi_{2}}$ can work for searching for the CME, we further study the stage evolution of $C_{\Psi_{2}}$, $C^{\perp}_{\Psi_{2}}$ and $R_{\Psi_{2}}$ for the four stages of heavy-ion collisions in the AMPT model. The results of original AMPT without the CME are presented in Fig.~\ref{fig-2}. We can see $C_{\Psi_{2}}$, $C^{\perp}_{\Psi_{2}}$ are flat at the initial stage, and then convex at the stage of after parton cascade. After the coalescence, $C_{\Psi_{2}}$ and $C^{\perp}_{\Psi_{2}}$ both trend to be flat, but they become more convex after hadronic rescatterings. However, as the ratio of $C_{\Psi_{2}}$ and $C^{\perp}_{\Psi_{2}}$, $R_{\Psi_{2}}$ is always flat and around the unit from initial stage to after hadronic rescatterings.

\begin{figure}
\includegraphics[scale=0.85]{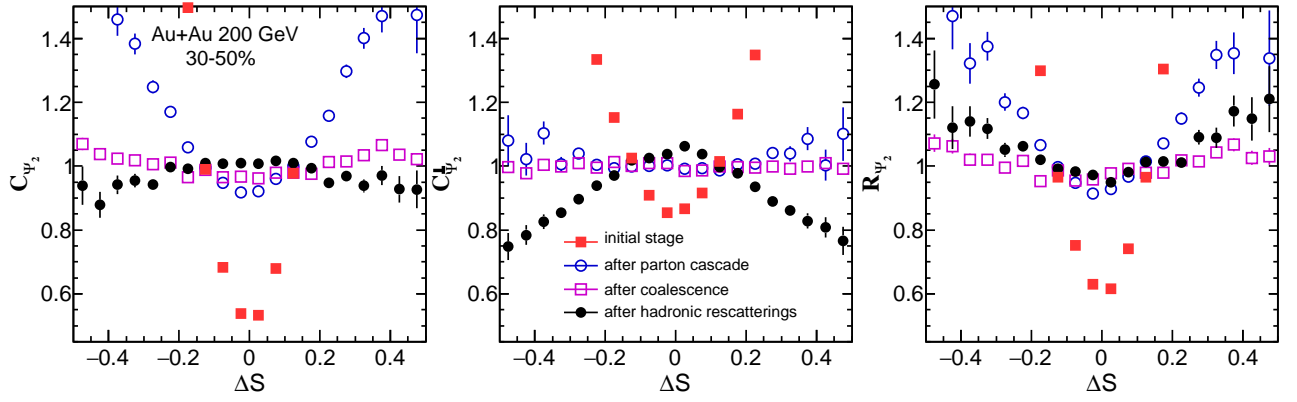}
\caption{(Color online)  $C_{\Psi_{2}}$, $C^{\perp}_{\Psi_{2}}$ and $R_{\Psi_{2}}$ in Au+Au collisions at 200 GeV (30-50\%) from different evolution stages of the AMPT model with a 10\% initial CME-induced charge separation.
}
\label{fig-3}
\end{figure}
At the same time,  we also calculated the stage evolution of $C_{\Psi_{2}}$, $C^{\perp}_{\Psi_{2}}$ and $R_{\Psi_{2}}$ for the AMPT model with the CME. As presented in Fig.~\ref{fig-3}, $C_{\Psi_{2}}$, $C^{\perp}_{\Psi_{2}}$ and $R_{\Psi_{2}}$ are most concave at the initial stage due to introducing the CME. Then after parton cascade, three results are still concave but the magnitude is weaken compared to that at initial stage, due to strong parton cascade. At the stage of after coalescence, three results trend to become flat. After hadronic rescatterings, $C_{\Psi_{2}}$ and $C^{\perp}_{\Psi_{2}}$ become convex while $R_{\Psi_{2}}$ becomes concave. In this way, the concave shape due to the CME survives from the final state interactions, which gives us a chance to search for the CME by using the new observable of $R_{\Psi_{2}}$. In the previous work, Ma et al.~\cite{Ma:2011uma} also investigated the evolution of $\gamma$ observable in the AMPT model which shows final state interactions strongly weaken the initial CME-induced charge separation. Our results indicates that the CME signal in $R_{\Psi_{2}}$ is suffers a similar fate to that in the $\gamma$ observable, i.e. the CME signal from the initial stage is weaken due to final state interactions~\cite{Ma:2011uma}. 

\begin{figure}
\includegraphics[scale=0.85]{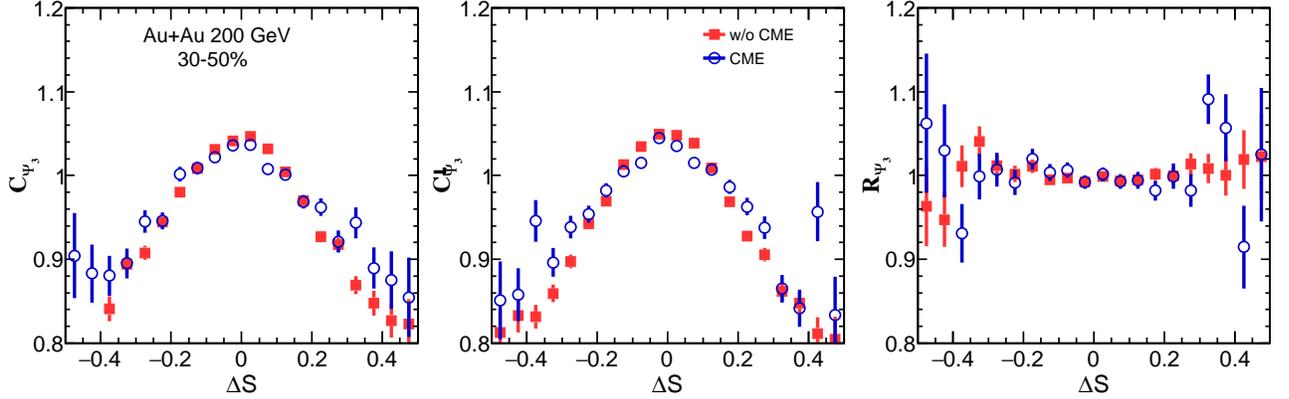}
\caption{(Color online)  $C_{\Psi_{3}}$, $C^{\perp}_{\Psi_{3}}$ and $R_{\Psi_{3}}$ in Au+Au collisions at 200 GeV (30-50\%) from the AMPT model without the CME and with a 10\% initial CME-induced charge separation.
}
\label{fig-4}
\end{figure}

\subsection{$C_{\Psi_{3}}$, $C^{\perp}_{\Psi_{3}}$ and $R_{\Psi_{3}}$}
\label{sec:part C}

We also study $R_{\Psi_{3}}$ which is defined to be respect to the third event plane $\Psi_{3}$. As the direction of magnetic field is expected to be not correlated to $\Psi_{3}$, some research~\cite{Magdy:2017yje} indicates that $R_{\Psi_{3}}$ from the background can not identify the CME signal and background. Therefore, we calculated $R_{\Psi_{3}}$ for the original AMPT model and the AMPT model with introducing the CME. The results are shown in Fig.~\ref{fig-4},  we can see that  $C_{\Psi_{3}}$ and $C^{\perp}_{\Psi_{3}}$ are convex, $R_{\Psi_{3}}$ are flat. Because the results from the original AMPT model is same as those from the AMPT model with the CME, which confirmes that $R_{\Psi_{3}}$ is indeed not sensitive to the CME.

\subsection{Sensitivity to the CME}

In previous work, Ma et al.~\cite{Ma:2011uma} have studied relationship between the traditional observable of $\gamma$ and the initial charge separation percentage due to the CME through the AMPT model, which indicates that $\gamma$ is not linearly response to the initial charge separation percentage if considering of final state interactions. It demonstrated that only when the charge separation percentage is large enough, e.g. more than 5\%, the effect on $\gamma$ from the CME can become visible. It is interesting to also study how sensitive to the CME the new observable of $R_{\Psi_{2}}$ is.

\begin{figure}
\includegraphics[scale=0.85]{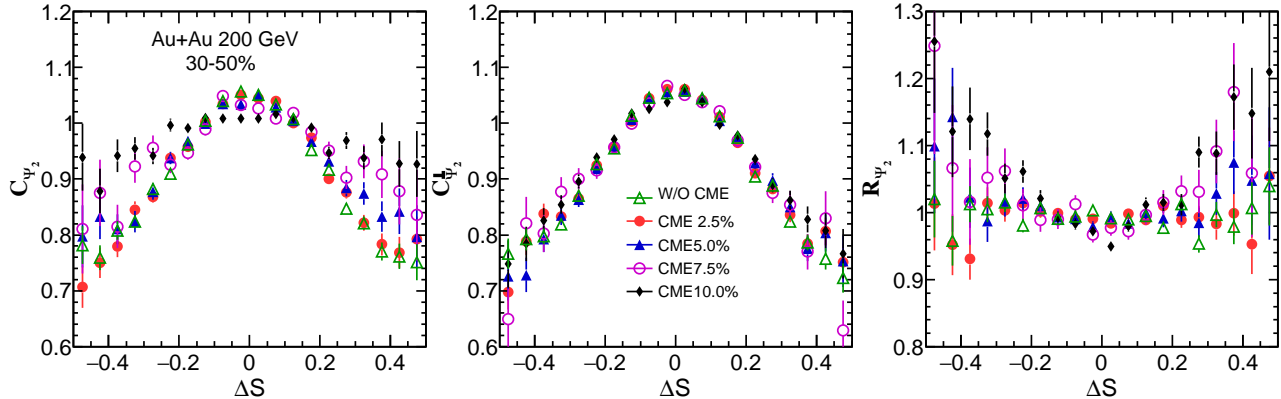}
\caption{(Color online)  $C_{\Psi_{2}}$, $C^{\perp}_{\Psi_{2}}$ and $R_{\Psi_{2}}$ in Au+Au collisions at 200 GeV (30-50\%) from the AMPT model without the CME and with different initial charge separation percentages.
}
\label{fig-5}
\end{figure}

Fig.~\ref{fig-5} shows the results of $C_{\Psi_{2}}$, $C^{\perp}_{\Psi_{2}}$ and $R_{\Psi_{2}}$ from the AMPT model with different initial charge separation percentages. The results from the original AMPT model without the CME is similar to those from the AMPT model with 2.5\% initial charge separation percentage, where $R_{\Psi_{2}}$ are both flat within the error bars. When introducing a 5\% initial charge separation percentage into the AMPT model, $C_{\Psi_{2}}$ become wider than the $C_{\Psi_{2}}$ with 2.5\% initial charge separation percentage, which makes $R_{\Psi_{2}}$ trend to be concave. With the initial charge separation percentage increases, the $C_{\Psi_{2}}$ becomes wider and wider, and concave $R_{\Psi_{2}}$ becomes narrower and narrower. Within our current event statistics (2 Million events for each case), our results show when the initial charge separation percentage is larger than 5\%, the shape of $R_{\Psi_{2}}$ starts to be sensitive to the CME. However, since real experiments have much more events than ours, it is possible for experimentalists to measure a even smaller percentage of CME signal based on the large experimental data sample. 

\begin{figure}
\includegraphics[scale=0.4]{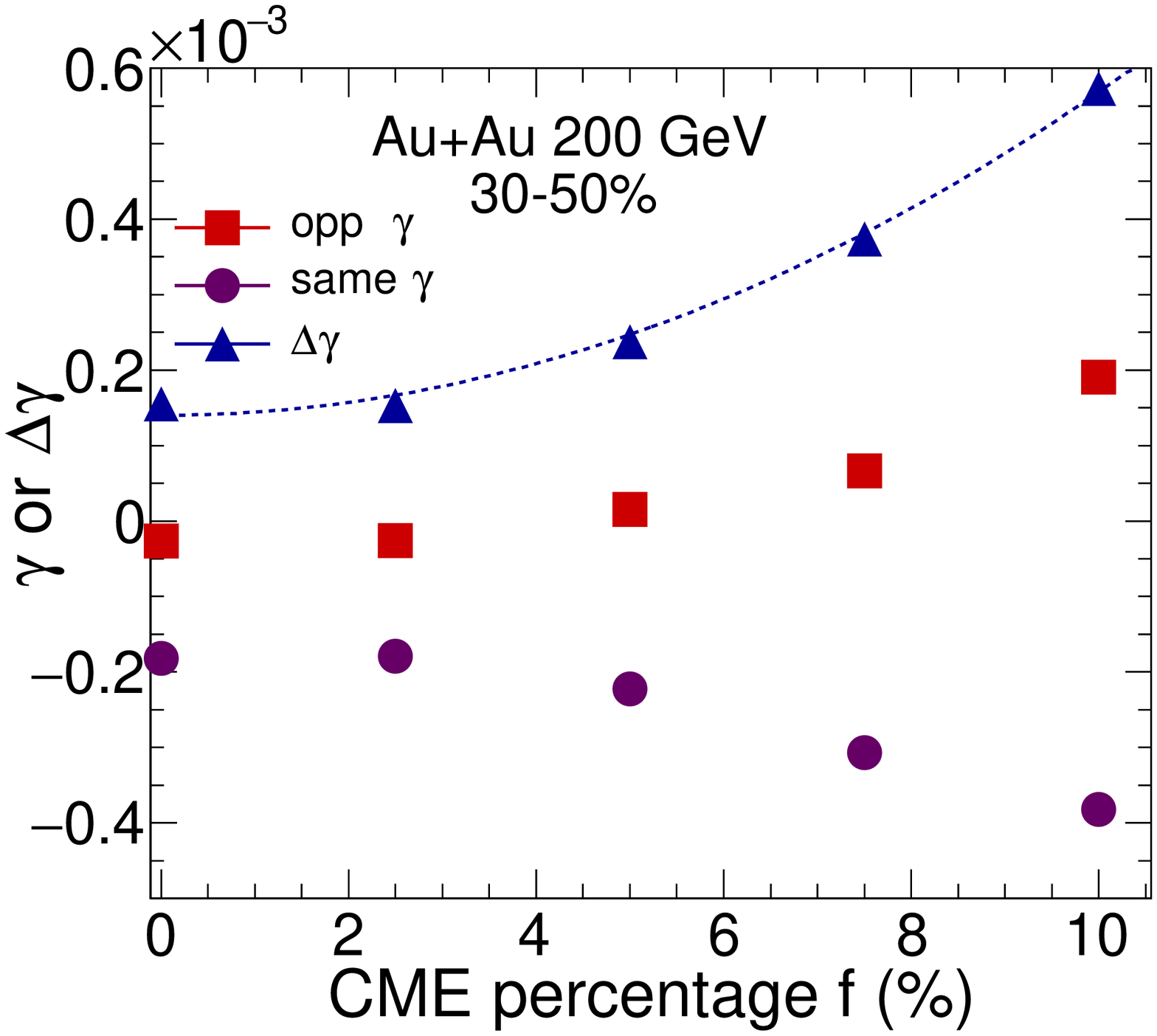}
\includegraphics[scale=0.4]{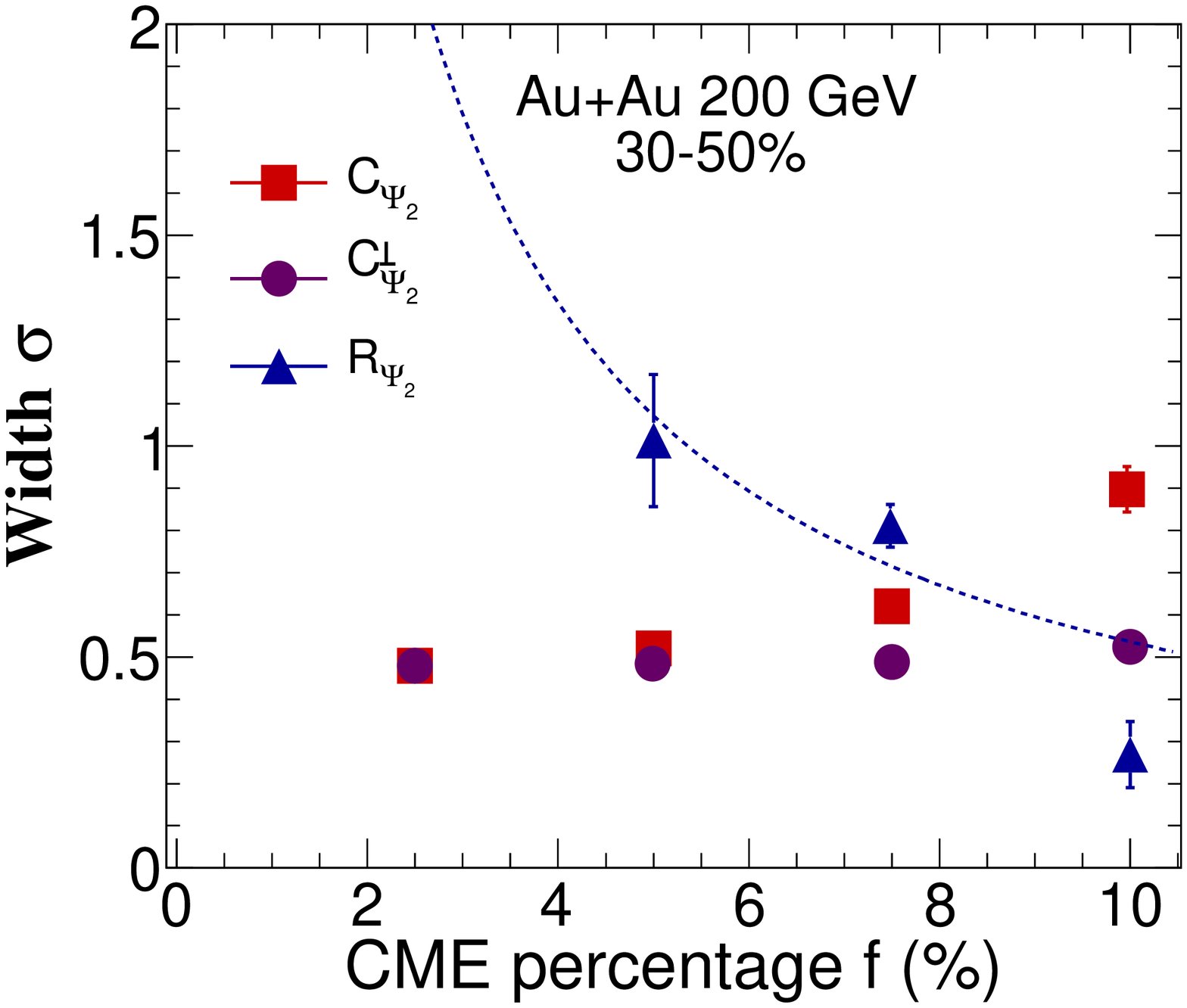}
\caption{(Color online)  The initial charge separation percentage dependence  of final $\gamma$, $\Delta\gamma$ (left plot), and the width of $C_{\Psi_{2}}$, $C^{\perp}_{\Psi_{2}}$ and $R_{\Psi_{2}}$ (right plot) in Au+Au collisions at 200 GeV (30-50\%), where the curves are fitting functions of $\Delta\gamma=Af^{2}+B$ and $\sigma=A/f$, respectively.
}
\label{fig-6}
\end{figure}

In order to compare the sensitivities to the CME between $\gamma$ and $R_{\Psi_{2}}$, we study how they depend on the initial charge separation percentage. In the left plot of Fig.~\ref{fig-6}, we show that the $\gamma$ and $\Delta\gamma$ have nonlinear responses to the initial charge separation percentage. The $\gamma$ and $\Delta\gamma$ from the AMPT with a 2.5\% initial charge separation percentage is almost same as those from the original AMPT model (0\%). $\gamma$ and $\Delta\gamma$ from the AMPT model with a 5.0\% initial charge separation percentage is slightly different from those with 0\% and 2.5\% initial charge separation percentages, which indicates it is difficult for using $\gamma$ to detect the CME if the initial charge separation percentage is less than 5.0\%. When the initial charge separation percentage increases from 5\% to 10\%, the $\gamma$ and $\Delta\gamma$ start to increase with the initial charge separation percentage, which is consistent with the previous results from Ma et al.~\cite{Ma:2011uma}.On the other hand, the right plot of Fig.~\ref{fig-6} shows the width $\sigma$ of $C_{\Psi_{2}}$, $C^{\perp}_{\Psi_{2}}$ and $R_{\Psi_{2}}$ distributions for different initial charge separation percentages in Au+Au collisions (30-50\%), where we apply a Gaussian function to fit the distributions of $C_{\Psi_{2}}$, $C^{\perp}_{\Psi_{2}}$ and $R_{\Psi_{2}}$.We can see that the width of $C_{\Psi_{2}}$ increases but $C^{\perp}_{\Psi_{2}}$ changes little, so the width of $R_{\Psi_{2}}$ decreases, when the initial charge separation percentage is larger than 5\%. Note that the width of $R_{\Psi_{2}}$ for 2.5\% is not plotted because the distribution of $R_{\Psi_{2}}$ for 2.5\% is so flat that we can not extract the width out by the fitting.

\begin{figure}
\includegraphics[scale=0.5]{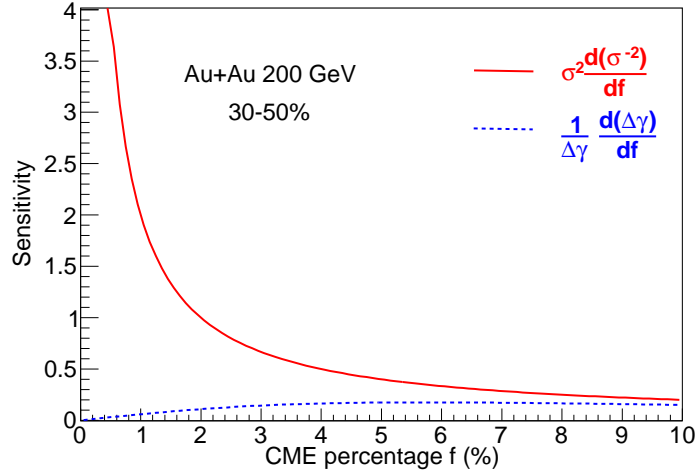}
\caption{(Color online)  The sensitivity to initial charge separation percentage dependence of $\Delta\gamma$ and $R_{\Psi_{2}}$  in Au+Au collisions at 200 GeV (30-50\%).
}
\label{fig-7}
\end{figure}

Because it is well known that $\Delta\gamma$ is proportional to $a_{1}^{2}$ (or $f^{2}$), we assume $\Delta\gamma=Af^{2}+B$ holds, where A and B are fitting parameters and B stands for the background contribution.  On the other hand, since the width $\sigma$ reflects the fluctuation of $\Delta$S, we simply assume it is inversely proportional to $f$ (or $a_{1}$), i.e. $\sigma=A/f$, to fit our result. The dash curves in the two panels of Fig.~\ref{fig-6} show our fitting functions to  $\Delta\gamma$  and the width $\sigma$, respectively. To further compare the sensitivities of the two observables to CME,  Fig.~\ref{fig-7} shows our defined sensitivities of (1/$\Delta\gamma$)(d$\Delta\gamma$/d$f$) and $\sigma^{2}$d($\sigma^{-2}$)/d$f$ as functions of the initial charge separation percentage, based on our fitting functions. Note that we choose $\sigma^{-2}$ instead of $\sigma$, because both $\sigma^{-2}$ and $\Delta\gamma$ are proportional to $f^{2}$, which makes the comparison more fair. We find that the sensitivity of $R_{\Psi_{2}}$ decreases but that of $\Delta\gamma$ increases with $f$. It shows that the width of $R_{\Psi_{2}}$ is more sensitive to the CME than $\Delta\gamma$ when the initial CME percentage is small. This is due to the sudden change of  curvature from a flat shape (without the CME) to a concave one (with the CME)  in terms of the shape of $R_{\Psi_{2}}$. However, when the initial CME percentage becomes large, two observables becomes similarly sensitive to the CME. Therefore, it is perferable to detect a small signal of the CME by using the new observable of $R_{\Psi_{2}}$, suppose ones have enough event statistics.

\section{Summary}
\label{sec:summary}
We have studied the chiral magnetic effect with the new observable of $R_{\Psi_{m}}$ within the framework of  a multiphase transport model without and with introducing CME-induced charge seperation. The results from mixing-particle method and shuffling-particle method are consistent with each other. We confirm that the shape of $R_{\Psi_{2}}$ distribution is flat for the background only, while it can be concave if with an amount of the CME, which reveals that $R_{\Psi_{2}}$ is capable of distinguishing the CME signal from the background. But for $R_{\Psi_{3}}$, it is not sensitive to the CME. We also presented the stage evolution of $R_{\Psi_{2}}$ distribution, which indicates the initial CME signal can be weakened by strong final state interactions, similarly as $\gamma$. We also compared the sensitivities to the CME between $R_{\Psi_{2}}$ and $\gamma$, and found that both observables show nonlinear responses to the CME. The shape of $R_{\Psi_{2}}$ show a larger sensitivity to the CME than $\gamma$ when the CME signal is small. However, measuring the shape of $R_{\Psi_{2}}$ for a small CME signal requires large event statistics.

\section*{ACKNOWLEDGMENTS}

We thank Roy Lacey and Jiangyong Jia for their valuable discussions. This work was supported by the National Natural Science Foundation of China under Grants No. 11890714, No. 11835002, No. 11421505, No. 11522547 and No. 11375251, the Key Research Program of the Chinese Academy of Sciences under Grant No. XDPB09.

\end{document}